\documentclass[11pt]{article}

\usepackage[T1]{fontenc}
\usepackage{textcomp}
\usepackage[centertags]{amsmath}
\usepackage[mediummath]{nccmath}
\usepackage{amsfonts}
\usepackage{amssymb}
\usepackage{mathtools}
\usepackage[pdftex]{hyperref}
\usepackage{graphicx}
\usepackage{graphbox}
\usepackage[numbers,sort&compress]{natbib}

\usepackage{paperinitial}

\paperinitialization{15mm}{15mm}{15mm}{15mm}{2pt}{10pt}

\DeclareMathOperator{\sgn}{sgn}
\DeclareMathOperator{\lcm}{lcm}

\newcommand{\vf}{\varphi}
\newcommand{\vk}{\varkappa}

\newcommand{\al}{\alpha}
\newcommand{\be}{\beta}
\newcommand{\ga}{\gamma}

\newcommand{\de}{\delta}
\newcommand{\De}{\Delta}

\newcommand{\la}{\lambda}

\newcommand{\N}{\mathbb{N}}
\newcommand{\Z}{\mathbb{Z}}

\begin{document}
\allowdisplaybreaks[4]
\frenchspacing

\title{{\Large\textbf{Three-frequency helical undulator as\\ a source of photons in composite twisted states}}}
%

\date{}

\author{%
O.V. Bogdanov${}^{1)}$\thanks{E-mail: \texttt{bov@tpu.ru}},\;
S.V. Bragin${}^{1)}$\thanks{E-mail: \texttt{svb38@tpu.ru}},\;
P.O. Kazinski${}^{1,2)}$\thanks{E-mail: \texttt{kpo@phys.tsu.ru}},\;
and
V.A. Ryakin${}^{1,2)}$\thanks{E-mail: \texttt{vlad.r.a.phys@yandex.ru}}\\[0.5em]
{\normalsize ${}^{1)}$Mathematics and Mathematical Physics Division,}\\
{\normalsize Tomsk Polytechnic University, Tomsk 634050, Russia}\\[0.5em]
{\normalsize ${}^{2)}$ Physics Faculty, Tomsk State University, Tomsk 634050, Russia}
}

\maketitle

\begin{abstract}

The properties of radiation from a three-frequency helical undulator are thoroughly investigated. It is shown that such undulators can be employed for generating photons in the so-called composite twisted states -- the states that are linear superpositions of modes with definite projections of the total angular momentum, amplitudes, relative phases, and polarizations. We find the explicit expressions and the selection rules for these parameters and establish that they can be governed in a predictable way by adjusting the parameters of the multifrequency helical undulator. In particular, the phases of three arbitrary modes admissible by selection rules in the composite state with definite energy can be made arbitrary by tuning the phases of one-frequency undulators comprising the three-frequency one. By solving Diophantine equations, we obtain simple expressions for the complex amplitude of coherent state of photons emitted by a three-frequency helical undulator and for the average number of radiated twisted photons in the case when the ratios of frequencies of the three-frequency undulator are rational numbers. The development of resonances and the control of composite states of radiated photons are studied numerically confirming the theoretical conclusions.

\end{abstract}

\section{Introduction}

Nowadays the photons prepared in twisted states are a useful tool for investigating excitations of rotational degrees of freedom of various quantum objects \cite{Nandi2004,Kapale2005,Afanas13,Mukherjee2018,Lange2022,QRTKrmp22,KazRyak23,Peshkov2023,KazSok2024,Lu2023,Kirschbaum2024,Liu2026, KKR_Rydberg2025} and their chirality \cite{Bazarov2025} (for these and many other applications of twisted photons see also \cite{AndBabAML,PadgOAM25,Roadmap16,KnyzSerb,ZWYB20}). By definition, the twisted states of photons are the quantum states with definite projection of the total angular momentum (TAM) on the given axis, projection of the momentum onto this axis, absolute value of the perpendicular momentum component, and helicity. Since the photons in such states carry the TAM projection that can be different from $\pm1$ and even be very large \cite{Chen2013,Fickler2016}, the new channels of quantum processes open or become dominating as against the same processes with photons prepared in plane-wave states \cite{Afanas13,Mukherjee2018,Lange2022,KazRyak23,Peshkov2023,SGACSSK,KazSok2024,Lu2023,Kirschbaum2024,Liu2026,KKR_Rydberg2025}. Coherent superpositions of twisted states with distinct projections of TAM provide a means for a coherent control of the quantum processes initiated by the photons prepared in the states with definite TAM projections. Such states have already found their applications in studying the interference effects in the Compton process \cite{Stock2015,Sherwin2017,Sherwin2020,Ivanov2022}, in manipulating cold atoms and vortices in Bose-Einstein condensates \cite{Franke-Arnold2007,Nandi2004,Kapale2005,MoxleyIII2016}, in increasing resolution in optical metrology \cite{Cheng2025}, and in a high-density information transfer \cite{Thide07,Anguita2014,Willner22,JiangWerner22,Chen2023,Shu2024,Sasaki2024,Kazinski2024,Manavalan2025,Aksenov2025}. In what follows, we call for brevity such states the composite twisted states or simply the composite states.

The experimental generation of photons in the composite states with tunable energy, spectrum with respect to the TAM projection, relative phases, and amplitudes is realized only in the optical range and below on the energy scale \cite{Willner22,Chen2023,Sasaki2024,Shu2024,Manavalan2025,Adamov2022,Gu2025,Tan2026,Zhang2019,Chen2019, Yue2017,Vasilyeu2009,Zheng2021,Shi2025}. In the optical and terahertz spectrum ranges, one usually employs ring-shaped phase plates with different topological charges \cite{Vasilyeu2009}, spatial light modulators \cite{Chen2013,Fickler2016,Anguita2014}, appropriately designed metasurfaces \cite{ZWYB20,Yue2017,Zhang2019,Chen2019,Zheng2021,Shi2025}, mode couplers in multi-mode fibers \cite{Gu2025}, or coherent beam combining \cite{Thide07,Chen2023,Shu2024,Sasaki2024,Kazinski2024,Aksenov2025,Adamov2022}. All these approaches are based on one or another type of conversion of the coherent plane-wave radiation to the twisted one and so they operate in a narrow bandwidth and have a relatively low brightness. We show in the preset paper (see also \cite{dfund}) that the $M$-frequency helical undulator directly generates photons in the composite states. It represents an adjustable bright source of photons prepared in such states in a wide range of photon energies. In particular, the $M$-frequency undulator provides a much brighter source of photons in the composite states in a much wider range of energies than coherent beam combining. Here some comments on the terminology are in order. We call a $M$-frequency undulator an undulator with the magnetic field being a linear superposition of the magnetic fields of $M$ coaxial one-frequency undulators with distinct frequencies. Then a $M$-frequency helical undulator is an undulator consisting of $M$ coaxial one-frequency helical undulators. The properties of radiation produced by multifrequency undulators have been studied for a long time starting with the paper \cite{Iracane1991} (see also \cite{Ciocci1993,Bazylev1993,Zhukovsky2021,Tripathi2011,Gajbhiye2023}). However, the spectrum of this radiation with respect to the projection of TAM and the explicit expressions for the radiation amplitudes in the general case of $M$-frequency undulator have been unknown until the paper \cite{dfund}. For the number of frequencies $M\geqslant5$, these amplitudes have not been obtained even in the basis of plane-wave photons \cite{Iracane1991,Ciocci1993,Bazylev1993,Zhukovsky2021,Tripathi2011,Gajbhiye2023,ZhukovskyFedorov2021,Mishra2020,Zhukovsky2017, Datolli2014,Mishra2009,Dattoli2006}.

In the present paper, we simplify the general formulas derived in \cite{dfund} and specialize them to three-frequency undulators. Then we consider in detail the case of a three-frequency helical undulator to demonstrate how such undulators can be employed for generation of photons in the composite states with preassigned properties. It turns out that the spectrum with respect to the TAM projection, the amplitudes of modes comprising the composite state of an emitted photon, and their relative phases can be adjusted by changing the parameters of the three-frequency helical undulator. For example, on enabling the properly chosen third frequency in the two-frequency helical undulator making it a three-frequency one, we can amplify the modes in the composite state of a radiated photon due to development of a resonance or generate a new mode with a certain projection of TAM. Changing the phases of one-frequency undulators constituting the multifrequency one, we can tune the relative phases of modes in the composite state. We also perform the numerical simulations of undulator radiation for the real-world parameters of undulators. These simulations corroborate the theoretical conclusions.

The paper is organized as follows. In Sec. \ref{General_Formulas}, we recapitulate and simplify the main formulas describing the dynamics of an electron in the magnetic field of $M$-frequency undulator and the one-particle radiation amplitude. In Sec. \ref{Rad_3_Freq_Undul}, we consider the special case of three-frequency undulator with rational ratios of frequencies. Section \ref{Helic_3Freq_Undul} is devoted to the properties of radiation generated by a three-frequency helical undulator. In Sec. \ref{Num_Simul}, the numerical simulations and the applications of the general theory are presented. In Conclusion, we summarize the results. Some calculations concerning the relative phases of modes in composite states are removed to Appendix \ref{Rel_Phases_Modes_App}. We use the notation and agreements adopted in the papers \cite{BKL2,KazRyakElUnd}. In particular, we use the system of units such that $\hbar=c=1$ and $e^2=4\pi\al$, where $\al$ is the fine structure constant. We also assume that $x\equiv x_1$, $y\equiv x_2$, and $z\equiv x_3$.

\section{Radiation from a $M$-frequency undulator}\label{General_Formulas}

For the convenience of the reader, we present in this section the general formulas describing the radiation of twisted photons by a $M$-frequency undulator obtained in the paper \cite{dfund}. Besides, we simplify the general formulas for the radiation amplitude derived in \cite{dfund}. In the next sections, we will consider the special case $M=3$ in detail.  Let the magnetic field of the undulator in the vicinity of the $z$ axis, along which the electron moves on average, be a superposition of the magnetic fields of $M$ coaxial one-frequency elliptical undulators (see, for example, \cite{BaKaStrbook,Bord.1,Zhukovsky2021,Dattoli2006})
\begin{equation}\label{magnetic_field}
    H_x=\sum_{i=1}^M H_x^i\sin \tilde{\vf}_i,\qquad H_y=\sum_{i=1}^M H_y^i\cos \tilde{\vf}_i,\qquad H_z=0,
\end{equation}
where $\tilde\vf_{i}=\pm2\pi z/l_i +\chi_{i}$, and $H_x^i$, $H_y^i$, and $\chi_i$  are some constants characterizing the magnetic field strength and the phase shifts of the one-frequency undulators relative to each other, $ l_i $ are the lengths of the sections of one-frequency undulators, which are assumed to be distinct for different $i$. For definiteness, we suppose that $l_1=\max l_i$. We refer to such a $M$-frequency undulator as comprised of $M$ one-frequency undulators. An ultrarelativistic electron with charge $e$ moves on average along the $z$ axis in the positive direction in the magnetic field \eqref{magnetic_field}. Then its trajectory is written as (see formula (5) in \cite{dfund})
\begin{equation}\label{trajectory}
\begin{gathered}
	x= x_0+ \sum_{i=1}^{M}a_{i} \cos\vf_i,\qquad y=y_0+ \sum_{i=1}^{M} b_i\sin\vf_i, \\
    z= z_0+\be_z
    t + \sum_{\substack{i,j=1,\\i< j}}^M
	\big[c_{ij}\sin(\vf_i+\vf_j)+d_{ij}\sin(\vf_i-\vf_j)\big] + \sum_{i=1}^M
	c_{ii}\sin(2\vf_i),
\end{gathered}
\end{equation}
where $\vf_{i}=\omega_{i} t +\chi_{i}$, $\omega_{i}=\pm2\pi\be_3/l_{i}$, and $x_0$, $y_0$, $z_0$  are some constants. In expression \eqref{trajectory}, the following notation has been introduced
\begin{equation}\label{a_b_be3_K}
\begin{gathered}
    a_i=\frac{eH_y^i}{m_e\ga\omega_i^2},\qquad b_i=-\frac{eH_x^i}{m_e\ga\omega_i^2},\\
    \be_z=1-\frac{1+K^2}{2\ga^2},\qquad K^2=\sum_{i=1}^M K_i^2,\qquad K_i^2=\frac{\omega_i^2\ga^2}{2}(a_i^2+b_i^2),
\end{gathered}
\end{equation}
where $m_e$ is the electron mass, $\ga$ is its Lorentz factor, $K_i$ is the undulator strength of the $i$-th undulator, and $K$ is the undulator strength of the $M$-frequency undulator. The amplitudes of longitudinal oscillations are given by
\begin{equation}\label{c_d}
	c_{ij}=\frac{\omega_i\omega_j}{2}\frac{a_ia_j-b_ib_j}{\omega_i+\omega_j},\qquad
    c_{ii}=\frac{\omega_i}{8}(a_i^2-b_i^2),\qquad
    d_{ij}=-\frac{\omega_i\omega_j}{2}\frac{a_ia_j+b_ib_j}{\omega_i-\omega_j},
\end{equation}
where it is supposed that $i<j$. The expression for the electron trajectory \eqref{trajectory} is valid under the assumption that $\ga\gg1$ and $K_{i}/\ga\ll1$. We assume that these estimates are satisfied. Furthermore, redefining the phases $\chi_i$ and the signs of $\omega_i$, one can always ensure that $a_i\geqslant0$, $b_i\geqslant0$. In what follows, we imply that these inequalities hold. Then the sign of $\omega_i$ specifies the helicity of the $i$-th elliptic undulator.

Let a $M$-frequency undulator have a length $L= Nl_1$, where $N\gg1$ is the number of sections of length $l_1$. We will neglect the contributions of edge radiation which is justified for sufficiently large $N$ in the photon energy range where the main radiation is concentrated. In this case, the expression for the trajectory \eqref{trajectory} is valid for $t\in[-TN/2,TN/2]$, where $TN:=L/\beta_z$. For the values of $t$ outside of this interval, we assume that the electron moves parallel to the $z$ axis with the velocity $\be_\parallel=\sqrt{1-1/\ga^2}$. Recall that the energy and, consequently, the Lorentz factor of the electron moving in a constant magnetic field are conserved.

Neglecting the contribution to radiation from the part of the electron trajectory for $t\not\in[-TN/2,TN/2] $, we obtain the average number of emitted twisted photons with helicity $s$, projection $m$ of the total angular momentum (TAM) onto the $z$ axis, projection of the momentum onto the same axis $k_z$, and modulus of the perpendicular component of momentum $k_\perp$ in the form \cite{dfund}
\begin{equation}\label{probab_I}
	dP(s,m,k_z,k_\perp)\approx e^2|\mathcal{A} |^2 n_\perp^3\frac{dk_zdk_\perp}{16\pi^2},
\end{equation}
where $ n^2_\perp = 1 - n_z^2 $, $ n_z = k_z/k_0 $, and $k_0$ is the photon energy. In the region where the main contribution to the radiation intensity is concentrated, $n_\perp\ga\lesssim\sqrt{1+K^2}$ and according to the general formula (38) of \cite{dfund} the complex conjugate one-particle amplitude of radiation of a twisted photon reads
\begin{equation}\label{totAmpl2}
\begin{split}
	\mathcal{A}^*\approx&\,  2\pi \sum_{\{n_i\},\{r_i\},\{p_{ij}\},\{q_{ij}\}=-\infty}^\infty
    \de_N\big(k_0(1 - n_z\be_z) -\sum_i n_i\omega_i \big) e^{ik_zz_0}
    j_{m-\tiny{\sum_i} (n_i+2r_i) + 2\tiny{\sum_{ij}} p_{ij}}^0\times\\
	&\times   \prod_{i=1}^{M} \Big[ J_{n_i+r_i-\tilde{q}_i-\tilde{p}_i}(\rho_i) J_{r_i}(\de_i) J_{p_{ii}}(\vk_{ii}) e^{in_i\chi_i}\Big]
    \prod_{\substack{k,l=1,\\k<l}}^M \Big[J_{q_{kl}}(\De_{kl}) J_{p_{kl}}(\vk_{kl}) \Big]\times\\
    &\times \Big\{ \be_z -
    \sum_{j=1}^{M}\frac{\omega_j}{k_\perp n_\perp} \Big(\frac{\rho_jJ_{n_j+r_j-\tilde{q}_j-\tilde{p}_j-s}(\rho_j)}{J_{n_j+r_j-\tilde{q}_j-\tilde{p}_j}(\rho_j)} -\frac{\de_jJ_{r_j-s}(\de_j)}{J_{r_j}(\de_j)} \Big) \Big\},
\end{split}
\end{equation}
where the summation is carried out over all the sets of the corresponding indices under the sum sign with indices $i<j$ for $q_{ij}$ and indices $i\leqslant j$ for $p_{ij}$. Furthermore,
\begin{equation}\label{tilde_q_tilde_p}
	\tilde{q}_i:=\sum_{j=i+1}^M q_{ij}- \sum_{j=1}^{i-1} q_{ji},\qquad \tilde{p}_i:=\sum_{j=i+1}^M p_{ij}+ \sum_{j=1}^{i-1} p_{ji},
\end{equation}
and
\begin{equation}\label{delta-shaped_seq}
	\de_N(x):=\frac{\sin(TNx/2)}{\pi x}.
\end{equation}
It is clear that
\begin{equation}
    \sum_{i=1}^M \tilde{q}_i=0,\qquad \sum_{i=1}^M\tilde{p}_i=2\sum_{\substack{i,j=1,\\i< j}}^M p_{ij}.
\end{equation}
In the expression for the amplitude \eqref{totAmpl2}, the notation has been introduced
\begin{equation}\label{j0_rho_delta_vk_Delta}
\begin{gathered}
	j^0_m:=j_m(k_\perp x_+^0,k_\perp x_-^0), \qquad x_\pm^0 := x_0 \pm i y_0, \qquad R_i := \frac{a_i + b_i}{2},
    \qquad D_i := \frac{a_i - b_i}{2}, \\
	\rho_i:=k_\perp R_i,\qquad \de_i:=k_\perp D_i,\qquad\vk_{ij}=k_3c_{ij}, \qquad \De_{ij}=k_3d_{ij},
\end{gathered}
\end{equation}
and the Bessel function $j_m(z,z^*)=\exp(im\arg z)J_m(|z|)$ has appeared (see \cite{BKL2} for the properties of this function). As compared to the general formula (38) of \cite{dfund}, the summation over the indices $p_{ij}$ and $q_{ij}$ with $i>j$ is absent in the expression for the radiation amplitude \eqref{totAmpl2} and the definitions of the quantities $c_{ij}$ and $d_{ij}$ with $i<j$ have been modified by a factor of two. In fact, in the amplitude \eqref{totAmpl2}, the summation has been performed over the aforementioned indices present in formula (38) of \cite{dfund}.

\section{Radiation from a three-frequency undulator}\label{Rad_3_Freq_Undul}

Let us now consider in more detail the case $M=3$. For large $N$, the radiation is concentrated near the harmonics
\begin{equation}\label{enSpect3}
	k_0=\frac{n_1\omega_1+ n_2\omega_2 +  n_3\omega_3}{1 - n_z\be_z}
    =:n_1\tilde{\omega}_1+n_2\tilde{\omega}_2 + n_3\tilde{\omega}_3,
\end{equation}
and
\begin{equation}\label{k_0_eta}
	k_0=\frac{\omega_1(\eta_1 n_1 + \eta_2 n_2 + \eta_3 n_3)}{1 - n_z\be_z},\qquad \eta_i:=\frac{\omega_i}{\omega_1},
\end{equation}
where $\eta_1=1$ and the integers $n_i$ are such that $k_0>0$.

Of particular interest is the special case where $\eta_i$ are rational numbers, $\eta_i=h_i/g_i$, $h_i\in \Z$, $g_i\in\N$. Then reducing the terms in parentheses in \eqref{k_0_eta} to a common denominator and factoring out the common factor, i.e., introducing
\begin{equation}
    g=\lcm(g_2,g_3),\qquad \tilde{g}_i:=g \eta_i,\qquad \mathrm{d}=\gcd(\tilde{g}_1,\tilde{g}_2, \tilde{g}_3),
    \qquad \la_i:=\tilde{g}_i/\mathrm{d},
\end{equation}
we obtain
\begin{equation}\label{omega_defn}
	k_0=\omega\frac{\la_1n_1 + \la_2 n_2 + \la_3 n_3}{1- n_z\be_z},\qquad \omega:=\omega_1\frac{\mathrm{d}}{g},\qquad
    \la_i=\frac{\omega_i}{\omega}.
\end{equation}
The numbers $\{\la_i\}$ are mutually prime and, therefore, according to the fundamental theorem on the greatest common divisor (see, for example, \cite{HasseBook}),
\begin{equation}\label{n_defn}
    \la_1n_1 + \la_2 n_2 + \la_3 n_3 =n,
\end{equation}
where $n$ is a nonzero integer with $\sgn n=\sgn\omega$. Hence
\begin{equation}\label{enSpect2_1}
    k_0=\frac{\omega n}{1 - n_z\be_z}\approx\frac{2\omega\ga^2n}{1+n_\perp^2\ga^2+K^2}.
\end{equation}
We shall call the number $n$ the principal quantum number or the harmonic number.

As can be seen from the formulas above, the harmonics defined by the relation \eqref{enSpect3} overlap in the case of rational $\eta_i$, i.e., there is an infinite number of triples $(n_1,n_2,n_3)$ corresponding to a given $n$. It is not difficult to describe the set of such triples. Let $\{n_i^0\}$, $i=\overline{1,3}$, be the B\'{e}zout coefficients for the set of mutually prime numbers $\{\la_i\}$. Then
\begin{equation}\label{n_i_sol}
    n_i= nn_i^0 +\de n_i,
\end{equation}
where $\{\de n_i\}$, $i=\overline{1,3}$, satisfy the linear homogeneous Diophantine equation
\begin{equation}\label{Diophantine_hom}
    \sum_{i=1}^3 \la_i \de n_i=0.
\end{equation}
There are various equivalent representations for the numbers $\{\de n_i\}$. In one of them (see \cite{Quinlan2022} and formula (69) of \cite{dfund})
\begin{equation}\label{de_n_i_sol_kab}
    \de n_1=\la_2 k_{21}+\la_3k_{31},\qquad \de n_2=-\la_1 k_{21}+\la_3 k_{32},\qquad \de n_3=-\la_1 k_{31} -\la_2 k_{32},
\end{equation}
where $k_{21}$, $k_{31}$, and $k_{32}$ are arbitrary integers. This representation is not minimal \cite{Quinlan2022} in the sense that there are different sets of the numbers $k_{21}$, $k_{31}$, and $k_{32}$ corresponding to the same values of $\{\de n_i\}$. In order to avoid this drawback, we also need another representation. In the paper \cite{Quinlan2022}, the algorithm was proposed for constructing a basis in the set of solutions to equation \eqref{Diophantine_hom} for an arbitrary finite number of unknowns. Here we employ a simpler method reducing equation \eqref{Diophantine_hom} to a linear Diophantine equation of two variables, which has an elementary representation for the general solution. Let
\begin{equation}
    \mathrm{d}_{23}=\gcd(\la_2,\la_3).
\end{equation}
Then according to the fundamental theorem on the greatest common divisor, the linear homogeneous Diophantine equation \eqref{Diophantine_hom} reduces to the system of equations
\begin{equation}\label{Diophant_systm}
    \la_1 \de n_1+ \mathrm{d}_{23} r=0,\qquad \la_2 \de n_2+\la_3 \de n_3= \mathrm{d}_{23} r,
\end{equation}
where $r\in \Z$. Since $\la_1$ and $\mathrm{d}_{23}$ are mutually prime, the first equation has a general solution
\begin{equation}
    \de n_1=-\mathrm{d}_{23} k_1,\qquad r=\la_1 k_1,
\end{equation}
where $k_1$ is an arbitrary integer. Introduce $\tilde{\la}_2:=\la_2/\mathrm{d}_{23}$, $\tilde{\la}_3:=\la_3/\mathrm{d}_{23}$, and the B\'{e}zout coefficients $(\tilde{n}^0_2,\tilde{n}^0_3)$ for $(\tilde{\la}_2,\tilde{\la}_3)$. Then the general solution to the second equation in \eqref{Diophant_systm} can be written as
\begin{equation}
    \de n_2=\tilde{n}^0_2 r-\tilde{\la}_3 k_2,\qquad \de n_3=\tilde{n}^0_3 r +\tilde{\la}_2 k_2,
\end{equation}
where $k_2$ is an arbitrary integer. As a result, we deduce the general solution to equation \eqref{Diophantine_hom} in the form
\begin{equation}\label{de_n_i_sol}
    \de n_1=-\mathrm{d}_{23} k_1,\qquad \de n_2=\tilde{n}^0_2 \la_1 k_1-\tilde{\la}_3 k_2,\qquad \de n_3=\tilde{n}^0_3 \la_1 k_1 +\tilde{\la}_2 k_2.
\end{equation}
It is clear that this representation is equivalent to \eqref{de_n_i_sol_kab}.

The representation \eqref{de_n_i_sol} allows one to obtain the explicit expressions for the amplitude of radiation of a twisted photon at a given harmonic $n$. It follows from formula \eqref{totAmpl2} that
\begin{equation}\label{totAmpl3}
\begin{split}
	\mathcal{A}^*\approx&\,  2\pi \sum_{n=\sgn(\omega)}^{\sgn(\omega)\infty}
    \de_N\big(k_0(1 - n_z\be_z) -n\omega \big) e^{ik_zz_0}\sum_{\substack{k_1,k_2,\{r_i\},\\ \{p_{ij}\},\{q_{ij}\}=-\infty}}^\infty
    j_{m-\tiny{\sum_i} (n_i+2r_i) + 2\tiny{\sum_{ij}} p_{ij}}^0\times\\
	&\times  \prod_{i=1}^{3} \Big[ J_{n_i+r_i-\tilde{q}_i-\tilde{p}_i}(\rho_i) J_{r_i}(\de_i) J_{p_{ii}}(\vk_{ii}) e^{in_i\chi_i}\Big]
    \prod_{\substack{k,l=1,\\k<l}}^3\Big[J_{q_{kl}}(\De_{kl}) J_{p_{kl}}(\vk_{kl}) \Big]\times\\
    &\times \Big\{ \be_z -
    \sum_{j=1}^{3}\frac{\omega_j}{k_\perp n_\perp} \Big(\frac{\rho_jJ_{n_j+r_j-\tilde{q}_j-\tilde{p}_j-s}(\rho_j)}{J_{n_j+r_j-\tilde{q}_j-\tilde{p}_j}(\rho_j)} -\frac{\de_jJ_{r_j-s}(\de_j)}{J_{r_j}(\de_j)} \Big) \Big\},
\end{split}
\end{equation}
where $\{n_i\}$ are expressed in terms of $k_{1,2}$ as in \eqref{n_i_sol} and \eqref{de_n_i_sol}.

\section{Three-frequency helical undulator}\label{Helic_3Freq_Undul}

Now we consider in detail the radiation of twisted photons by electrons propagating in a three-frequency helical undulator. In such an undulator, the electron moves along the trajectory \eqref{trajectory} with
\begin{equation}\label{a_b_helical}
	a_1 = b_1 = r_1, \qquad a_2 = b_2 = r_2,  \qquad a_3 = b_3 = r_3,
\end{equation}
and the magnetic field \eqref{magnetic_field} in the vicinity of the electron trajectory is equal to
\begin{equation}\label{magnetic_field_3freq_hel}
    H_x=H_x^1\sin \tilde{\vf}_1 + H_x^2\sin \tilde{\vf}_2 +H_x^3\sin \tilde{\vf}_3,\qquad H_y=- H_x^1\cos \tilde{\vf}_1 -H_x^2\cos \tilde{\vf}_2 - H_x^3\cos \tilde{\vf}_3,\qquad H_z=0.
\end{equation}
This magnetic field is a linear superposition of the magnetic fields of three one-frequency helical undulators. In general, we call a multifrequency helical undulator an undulator with the magnetic field being a linear superposition of the magnetic fields of coaxial one-frequency helical undulators with distinct frequencies. It follows from formulas \eqref{c_d}, \eqref{j0_rho_delta_vk_Delta}, and \eqref{a_b_helical} that the quantities entering into the arguments of the Bessel functions in the complex conjugate amplitude \eqref{totAmpl2} are written as
\begin{equation}
	R_{i} = r_{i} , \qquad D_{i}= \de_{i}  = 0,\qquad d_{ij}=-\frac{\omega_ir_i \omega_jr_j}{\omega_i-\omega_j},
    \qquad c_{ij}=\vk_{ij}=0, \qquad i,j \in \{1,2,3\}.
\end{equation}
The electron trajectory \eqref{trajectory} turns into
\begin{equation}
	x= x_0+ \sum_{i=1}^{3} r_{i} \cos\vf_i,\qquad y=y_0+ \sum_{i=1}^{3} r_i\sin\vf_i, \qquad
    z= z_0+\be_z
    t + \sum_{\substack{i,j=1,\\i< j}}^3
	d_{ij}\sin(\vf_i-\vf_j),
\end{equation}
and
\begin{equation}
    K_i=\ga|\omega_i| r_i.
\end{equation}
Since $J_n(0)=\de_{n0}$, all the Bessel functions having the arguments $\de_{i}$ or $\vk_{ij}$ are replaced by the Kronecker symbols. As a result, the complex conjugate one-particle amplitude of radiation of a twisted photon \eqref{totAmpl2} is reduced to
\begin{equation}\label{3omegaAmpl2}
\begin{split}
	\mathcal{A}^* \approx &\,  2\pi \sum_{n_1,n_2, n_3=-\infty}^\infty
	\de_N\big(k_0(1 -n_z\be_z)- n_1\omega_1 -n_2\omega_2 - n_3\omega_3 \big)
	j_{m- n_1 -n_2 -n_3}^0 e^{ik_zz_0+in_1\chi_1+in_2\chi_2 + in_3\chi_3} \times\\
    &\times  \Big[ \be_z J_{n_1n_2n_3}-
    \frac{\omega_1\rho_1}{k_\perp n_\perp} J_{n_1-s,n_2,n_3} -\frac{\omega_2\rho_2}{k_\perp n_\perp} J_{n_1,n_2-s,n_3} -\frac{\omega_3\rho_3}{k_\perp n_\perp} J_{n_1,n_2,n_3-s}\Big],
\end{split}
\end{equation}
where the special function has been introduced
\begin{equation}\label{6J_function}
\begin{split}
    J_{n_1n_2n_3}(\rho_1,\rho_2,\rho_3;\De_{12},\De_{13},\De_{23}):=&\sum_{q_{12},q_{13},q_{23}  =-\infty}^\infty J_{n_1-q_{12}-q_{13}}(\rho_1) J_{n_2+q_{12}-q_{23}}(\rho_2)J_{n_3+q_{13}+q_{23}}(\rho_3)\times\\
    &\times J_{q_{12}}(\De_{12})J_{q_{13}}(\De_{13})J_{q_{23}}(\De_{23}).
\end{split}
\end{equation}
For brevity, the arguments of this special function have been omitted in \eqref{3omegaAmpl2}. The factors at these special functions in \eqref{3omegaAmpl2} can be written as
\begin{equation}
    \frac{\omega_i\rho_i}{k_\perp n_\perp}= \frac{\omega_i r_i}{n_\perp}=\sgn(\omega_i)\frac{K_i}{n_\perp\ga}.
\end{equation}
In the case when $\{\eta_i\}$ are rational numbers, the complex conjugate one-particle amplitude of radiation of a twisted photon \eqref{totAmpl3} becomes
\begin{equation}\label{3omegaAmpl2_rational}
\begin{split}
	\mathcal{A}^* \approx &\,  2\pi \sum_{n=\sgn(\omega)}^{\sgn(\omega)\infty}
	\de_N\big(k_0(1-n_z\be_z)- n\omega \big)e^{ik_3z_0} \sum_{k_1,k_2=-\infty}^\infty
	j_{m- n_1 -n_2 -n_3}^0 e^{in_1\chi_1+in_2\chi_2 + in_3\chi_3} \times\\
    &\times  \Big[ \be_z J_{n_1n_2n_3}-
    \frac{\omega_1\rho_1}{k_\perp n_\perp} J_{n_1-s,n_2,n_3} -\frac{\omega_2\rho_2}{k_\perp n_\perp} J_{n_1,n_2-s,n_3} -\frac{\omega_3\rho_3}{k_\perp n_\perp} J_{n_1,n_2,n_3-s}\Big].
\end{split}
\end{equation}
where $\{n_i\}$ are expressed through $k_1$ and $k_2$ as it is given in \eqref{n_i_sol} and \eqref{de_n_i_sol}, and the fundamental frequency $\omega$ is defined in \eqref{omega_defn}.

As long as the Bessel function, $J_n(x)$, tends rapidly to zero in increasing $|n|$ in the domain $|n|>|x|$, the summation over $\{q_{ij}\}$ in the definition of the special function \eqref{6J_function} runs effectively in the range specified by the inequalities
\begin{equation}\label{ineq_rho_delta}
    -\rho_i\lesssim n_i-\tilde{q}_i\lesssim\rho_i,\qquad -|\De_{ij}|\lesssim q_{ij}\lesssim |\De_{ij}|,
\end{equation}
where
\begin{equation}\label{rho_delta}
    \rho_i\approx\Big|\frac{n}{\la_i}\Big|\frac{K_i}{K}\frac{2n_\perp\ga K}{1+n_\perp^2\ga^2+K^2},\qquad \De_{ij}\approx-\frac{2n}{\la_i-\la_j} \frac{K_i K_j}{1+n_\perp^2\ga^2+K^2}.
\end{equation}
The approximate expression for the radiation spectrum \eqref{enSpect2_1} has been taken into account in the expressions for  $\rho_i$ and $\De_{ij}$. The inequalities \eqref{ineq_rho_delta} also imply that the special function \eqref{6J_function} is essentially different from zero when
\begin{equation}\label{ineq_n_i}
    -\sum_{i=1}^3\rho_i\lesssim \sum_{i=1}^3 n_i\lesssim \sum_{i=1}^3\rho_i,\qquad -\rho_i -\sum_{j=i+1}^3 |\De_{ij}|- \sum_{j=1}^{i-1} |\De_{ji}| \lesssim  n_i\lesssim \rho_i +\sum_{j=i+1}^3 |\De_{ij}|+ \sum_{j=1}^{i-1} |\De_{ji}|.
\end{equation}
These inequalities determine the range of parameters $\{n_i\}$ where the absolute value of the special function \eqref{6J_function} is non-negligible. In this region, we also have to single out the subset obeying the constraint \eqref{n_defn} for a given harmonic. As is seen from \eqref{n_defn} and \eqref{rho_delta}, the resulting domain of parameters $\{n_i\}$ in the corresponding three-dimensional space stretches proportionally to $|n|$ as $|n|$ increases.

Because $j_m(0,0)=\de_{m0}$, the selection rule arises in the radiation amplitude \eqref{3omegaAmpl2},
\begin{equation}\label{sel_rul3}
    m = n_1+n_2 + n_3,
\end{equation}
for $ k_\perp |x_+^0| \ll 1 $, which is a generalization of the selection rule for radiation of twisted photons by a one-frequency undulator \cite{SasMcNu,HemMarRos11,HemMar12,HKDXMHR,BHKMSS,Rubic17,BKL2,BKL4}. This selection rule and the energy spectrum \eqref{enSpect3} possess a simple interpretation in terms of photons radiated by the electron and the virtual photons mediating between the electron and the undulator \cite{dfund}. In moving in the electromagnetic field of the undulator, the electron absorbs from (for $n_i\omega_i>0$) or gives to (for $n_i\omega_i<0$) the undulator field $|n_i|$ virtual photons with the frequencies $|\omega_i|$, $i=\overline{1,3}$, and also radiates a single physical photon with the frequency \eqref{enSpect3}. The virtual photons have the helicity $\sgn\omega_i$. Since $\sgn n_i=\sgn \omega_i$ when the electron absorbs a virtual photon created by the undulator and $\sgn n_i=-\sgn\omega_i$ when the undulator absorbs a virtual photon delivered by the electron, the selection rule \eqref{sel_rul3} just manifests the conservation of the TAM projection onto the $z$ axis, i.e., the conservation of the projection of TAM absorbed by the electron and transferred to a single radiated physical photon.

\begin{figure}[t!]
\centering
\includegraphics*[width=0.98\linewidth]{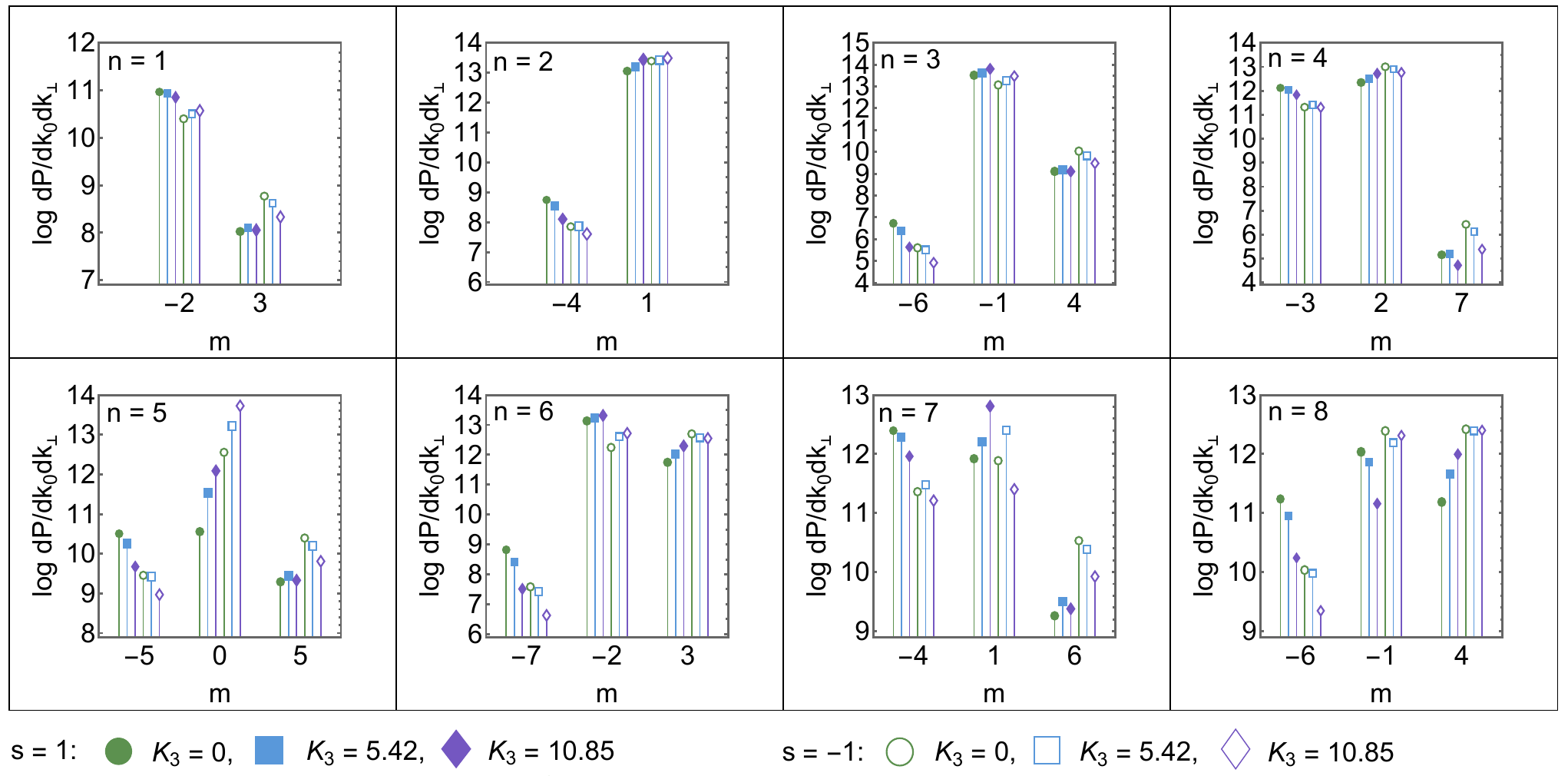}
\caption{{\footnotesize The spectrum of the TAM projections of radiation at the first eight harmonics of the three-frequency helical undulator at $n_\perp =K/\gamma$ and different $K_3$ and $s$. The energy of electrons in the undulator is $8$ GeV, the lengths of sections $(l_1,l_2,l_3)=(1,-3/2,7/2)\times 35$ cm. The magnetic field strengths are $(H^1_{x,y},H^2_{x,y},H^3_{x,y})=(0.2,1.16,1.16)\times 10^4$ G for the undulator strength parameters $(K_1,K_2,K_3)=(6.54,8.44,10.85)$. The energy of photons at the fundamental harmonic $k_0=1.87$ eV for these parameters. The undulator phases $\chi_i=0$. Only those values of the TAM projections are shown that give non-negligible contributions. It is seen that the selection rule \eqref{sel_rul3_part} is fulfilled.}}
\label{Plot_Undul_23m7_m_distr}
\end{figure}

For $k_\perp|x_+^0|\ll1$ and rational $\{\eta_i\}$, the spectrum with respect to the projection of TAM at a given harmonic $n$ has the form
\begin{equation}\label{sel_rul3-0}
	m=n( n_1^0 + n_2^0 + n_3^0 ) + \la_{21}k_{21} + \la_{31}k_{31} + \la_{32} k_{32},
\end{equation}
where formulas \eqref{n_i_sol}, \eqref{de_n_i_sol_kab}, and \eqref{sel_rul3} have been employed. The relation \eqref{sel_rul3-0} is equivalent to
\begin{equation}\label{sel_rul3-1}
	m=n( n_1^0 + n_2^0 + n_3^0 ) + \la_{21}\tilde{k}_{21} + \la_{31}\tilde{k}_{31}
    = n( n_1^0 + n_2^0 + n_3^0 ) +\big(\la_{12}\tilde{n}^0_2 +\la_{13}\tilde{n}^0_3 \big)k_1 +\tilde{\la}_{23}k_2,
\end{equation}
where $\tilde{k}_{21}$ and $\tilde{k}_{31}$ are arbitrary integer numbers. In the equality \eqref{sel_rul3-1}, we have used the representation \eqref{n_i_sol}, \eqref{de_n_i_sol} and, for conciseness, introduced the notation $\la_{21}:=\la_2-\la_1$, $\la_{31}=\la_3-\la_1$, $\la_{32}:=\la_3-\la_2$ and so on. It follows from the first representation in \eqref{sel_rul3-1} that any integer value of $m$ can be realized at a given harmonic for mutually prime $\la_{21}$ and $\la_{31}$. If $\la_{21}$ and $\la_{31}$ have a nontrivial greatest common divisor, then the selection rule is fulfilled
\begin{equation}\label{sel_rul3-2}
    m=n( n_1^0 + n_2^0 + n_3^0 ) + \mathrm{d}_{123} r, \qquad r\in \Z,
\end{equation}
where
\begin{equation}
    \mathrm{d}_{123}:=\gcd(\la_{21},\la_{31}).
\end{equation}
It means that the spectrum of admissible projections of TAM at a given harmonic $n$ is equidistant (see, e.g., Fig. \ref{Plot_Undul_23m7_m_distr}). The fundamental theorem on the greatest common divisor and the relations \eqref{sel_rul3-0} and \eqref{sel_rul3-1} imply that for mutually prime $\{\la_i\}$:
\begin{equation}
    \mathrm{d}_{123}\equiv \gcd(\la_{21},\la_{31})=\gcd(\la_{21},\la_{31},\la_{32})=\gcd(\tilde{\la}_{23},\la_{12}\tilde{n}^0_2 +\la_{13}\tilde{n}^0_3).
\end{equation}
Notice that the selection rule \eqref{sel_rul3-2} specifies only the admissible values of $m$. The magnitude of contributions with distinct $m$ allowed by the selection rule \eqref{sel_rul3-2} can be very different, including almost vanishing. The analysis carried out in the paper \cite{dfund} reveals that, among all the sets $\{n_i\}$ corresponding to the principal quantum number $n$, the considerable contribution comes from only those sets where the number of virtual photons, $|n_i|$, is not larger than the harmonic number $|n|$ by the order of magnitude. Furthermore, among such $n_i$, those give the leading contributions that correspond to the virtual photons with lesser energies $|\omega_i|$ and larger undulator strengths $K_i$. This property allows one to adjust the parameters of the multifrequency undulator so it generates the photons in composite states with various spectra with respect to the TAM projection. Moreover, it enables one to change the complex amplitudes of the modes comprising the composite state of a radiated photon. The first inequality in \eqref{ineq_n_i} bounds the achievable maximum value of $|m|$ of radiated photons. The examples of three-frequency undulators generating photons in diverse composite states are considered in Sec. \ref{Num_Simul}.

Let us obtain the amplitude of a coherent state of photons radiated at a given harmonic $n$ for $k_\perp|x_+^0|\ll1$ and rational $\{\eta_i\}$. In this case, the summation with respect to $k_1$ and $k_2$ in \eqref{3omegaAmpl2_rational} is performed over only those terms that obey the selection rule \eqref{sel_rul3}. It follows from the second equality in \eqref{sel_rul3-1} that for given $m$ and $n$,
\begin{equation}\label{k_12_sol}
    k_1=k_1^0+\frac{\tilde{\la}_{23}}{\mathrm{d}_{123}}k,\qquad  k_2=k_2^0-\frac{\la_{12}\tilde{n}^0_2 +\la_{13}\tilde{n}^0_3 }{\mathrm{d}_{123}}k,\qquad k\in\Z,
\end{equation}
where
\begin{equation}\label{k01_k02}
     k_1^0=\frac{m-n( n_1^0 + n_2^0 + n_3^0 )}{\mathrm{d}_{123}} b_1^0,\qquad k_2^0=\frac{m-n( n_1^0 + n_2^0 + n_3^0 )}{\mathrm{d}_{123}} b_2^0,
\end{equation}
and $b^0_1$, $b^0_2$ are the B\'{e}zout coefficients for
\begin{equation}
    \frac{\la_{12}\tilde{n}^0_2 +\la_{13}\tilde{n}^0_3}{\mathrm{d}_{123}},\qquad \frac{\tilde{\la}_{23}}{\mathrm{d}_{123}},
\end{equation}
respectively. It is assumed in the solution \eqref{k01_k02} that the fraction appearing in the expressions for $k^0_1$ and $k^0_2$ at the B\'{e}zout coefficients is an integer number, which is a necessary condition for solvability of \eqref{sel_rul3-1} for given $m$ and $n$. Substituting \eqref{k_12_sol} into \eqref{n_i_sol} and \eqref{de_n_i_sol}, we arrive at the representation
\begin{equation}\label{n_i_sol_m}
    n_1=\bar{n}_1+\frac{\la_{32}}{\mathrm{d}_{123}}k, \qquad n_2=\bar{n}_2+\frac{\la_{13}}{\mathrm{d}_{123}}k,\qquad n_3=\bar{n}_3+\frac{\la_{21}}{\mathrm{d}_{123}}k,
\end{equation}
where
\begin{equation}\label{bar_n_i}
    \bar{n}_1=nn_1^0-\mathrm{d}_{23}k_1^0,\qquad \bar{n}_2=nn_2^0 +\tilde{n}_2^0\la_1k_1^0-\tilde{\la}_3 k_2^0,\qquad \bar{n}_3=nn_3^0 +\tilde{n}_3^0\la_1k_1^0 +\tilde{\la}_2 k_2^0.
\end{equation}
By construction,
\begin{equation}\label{bar_n_restr}
    \sum_{i=1}^3\la_i\bar{n}_i=n,\qquad\sum_{i=1}^3\bar{n}_i=m.
\end{equation}
Then the complex conjugate one-particle amplitude of radiation of a twisted photon with the given projection of TAM $m$ turns into
\begin{equation}\label{amplitude_rad_m}
\begin{split}
	\mathcal{A}^* = &\,  2\pi \sum_{n=\sgn(\omega)}^{\sgn(\omega)\infty}
	\de_N\big(k_0(1 - n_z\be_z)- n\omega \big)e^{ik_zz_0 +i\tiny{\sum_i} \bar{n}_i\chi_i}  \times\\
    &\times  \Big[ \be_z \tilde{J}_{\bar{n}_1\bar{n}_2\bar{n}_3}-
    \frac{\omega_1\rho_1}{k_\perp n_\perp} \tilde{J}_{\bar{n}_1-s,\bar{n}_2,\bar{n}_3} -\frac{\omega_2\rho_2}{k_\perp n_\perp} \tilde{J}_{\bar{n}_1,\bar{n}_2-s,\bar{n}_3} -\frac{\omega_3\rho_3}{k_\perp n_\perp} \tilde{J}_{\bar{n}_1,\bar{n}_2,\bar{n}_3-s}\Big],
\end{split}
\end{equation}
where
\begin{multline}
    \tilde{J}_{n_1n_2n_3}(\rho_1,\rho_2,\rho_3;\De_{12},\De_{13},\De_{23};\chi_1,\chi_2,\chi_3):=\sum_{k=-\infty}^\infty e^{i(\la_{32} \chi_1+\la_{13}\chi_2 + \la_{21}\chi_3)k/\mathrm{d}_{123}}\times\\
    \times J_{n_1+\la_{23}k/\mathrm{d}_{123},n_2+\la_{13}k/\mathrm{d}_{123},n_3+\la_{21}k/\mathrm{d}_{123}}(\rho_1,\rho_2,\rho_3;\De_{12},\De_{13},\De_{23}).
\end{multline}
This amplitude is put to zero for those $m$ that do not satisfy the selection rule \eqref{bar_n_restr}. The complex conjugate of the expression \eqref{amplitude_rad_m} coincides up to a normalization factor with the complex amplitude of the coherent state of photons radiated by the electrons in the undulator (see, e.g., \cite{BKL2,Glaub2,KlauSud}). Discarding the interference contributions between the terms with distinct $n$, the average number of twisted photons with the projection of TAM $m$ emitted by the electron in the undulator reads
\begin{multline}\label{average_num_tw_phot}
	dP(s,m,k_3,k_\perp) =   \al\pi \sum_{n=\sgn(\omega)}^{\sgn(\omega)\infty}
	\de_N^2\big(k_0(1 - n_z\be_z)- n\omega \big)
    \times\\
    \times \Big|  \be_z \tilde{J}_{\bar{n}_1\bar{n}_2\bar{n}_3}-
    \frac{\omega_1\rho_1}{k_\perp n_\perp} \tilde{J}_{\bar{n}_1-s,\bar{n}_2,\bar{n}_3} -\frac{\omega_2\rho_2}{k_\perp n_\perp} \tilde{J}_{\bar{n}_1,\bar{n}_2-s,\bar{n}_3} -\frac{\omega_3\rho_3}{k_\perp n_\perp} \tilde{J}_{\bar{n}_1,\bar{n}_2,\bar{n}_3-s} \Big|^2n_\perp^3 dk_z dk_\perp,
\end{multline}
where it is supposed that this expression vanishes when the selection rule with respect to the projection of TAM (the second equality in \eqref{bar_n_restr}) does not hold.

The explicit expression \eqref{amplitude_rad_m} for the complex conjugate amplitude of radiation of a twisted photon makes it possible to obtain the relative phases of the modes with distinct projections of TAM in the composite state of a radiated photon at a given harmonic $n$. As long as there are three arbitrary parameters $\{\chi_i\}$, the relative phases of any three modes with different $m$ admitted by the selection rules \eqref{bar_n_restr} can be tuned to any preassigned values due to the phase factor
\begin{equation}
    e^{i\tiny{\sum_i} \bar{n}_i\chi_i},
\end{equation}
in the amplitude \eqref{amplitude_rad_m}. Therefore, the three-frequency helical undulators can produce the photons in the composite state with preassigned projections of TAM, amplitudes, and relative phases of any three modes in this state. In general, the number of modes in the composite state of a radiated photon with adjustable complex amplitudes is equal to the number of distinct frequencies of the helical undulator $M$. Some examples of the parameters of three-frequency helical undulators generating the photons in such states are presented in Sec. \ref{Num_Simul}.


In a general case, the relative phases of modes in the composite state of photons radiated by the undulator depends on the quantum numbers $s$ and $n_\perp$. One can get rid of this dependence if one takes all the phases $\chi_i$ equal to each other, i.e., $\chi_i=\bar{\chi}$ (a more general case leading to the same result is considered in Appendix \ref{Rel_Phases_Modes_App}). Then the complex conjugate one-particle amplitude of radiation of a twisted photon with the projection of TAM $m$ becomes
\begin{equation}\label{amplitude3_rad_m}
\begin{split}
	\mathcal{A}^* = &\,  2\pi \sum_{n=\sgn(\omega)}^{\sgn(\omega)\infty}
	\de_N\big(k_0(1 - n_z\be_z)- n\omega \big)e^{ik_zz_0 +i\bar{\chi}m}
    \times\\
    &\times \sum_{k=-\infty}^\infty \Big[ \be_z J_{n_1n_2n_3}-
    \frac{\omega_1\rho_1}{k_\perp n_\perp} J_{n_1-s,n_2,n_3} -\frac{\omega_2\rho_2}{k_\perp n_\perp} J_{n_1,n_2-s,n_3} -\frac{\omega_3\rho_3}{k_\perp n_\perp} J_{n_1,n_2,n_3-s}\Big],
\end{split}
\end{equation}
where $\{n_i\}$ are expressed through $k$ as in \eqref{n_i_sol_m}. It is seen from formula \eqref{amplitude3_rad_m} that the relative phases of modes in the composite state of a radiated photon are $\bar{\chi}m$ or $\bar{\chi}m+\pi$, i.e., they depend linearly on $m$ and can be made arbitrary for any two prescribed modes allowed by the selection rule. The average number of twisted photons radiated by such an undulator is given by the expression \eqref{average_num_tw_phot}, where one must set $\chi_{1,2,3}=0$ in all the functions $\tilde{J}_{\tilde{n}_1\tilde{n}_2\tilde{n}_3}$.

\section{Numerical simulations}\label{Num_Simul}

As it has been shown in Sec. \ref{Helic_3Freq_Undul}, the multifrequency undulators being the compositions of one-frequency helical undulators provide a means for generation of photons in composite states. The parameters of these states are governed in a rather simple way by changing the parameters of the multifrequency undulator -- the magnitudes of the magnetic fields, $H^i_{x,y}$, with distinct periods and the phases $\chi_i$. In particular, the turning on of the additional frequency in the multifrequency undulator can produce the new modes in the composite states of radiated photons with preassigned projections of TAM, amplitudes, and phases. Furthermore, the inclusion of the additional frequency allows one to amplify the existing modes in the composite state of photons radiated by the undulator with a lesser number of working frequencies. In the last case, in fact, the resonance appears at certain radiation harmonics and projections of TAM provided the activated frequency is properly chosen.

In order to produce the resonance at a given harmonic $n$ with a certain value of the TAM projection $m$, it is necessary that the turning on of the new frequency $M$ in the $(M-1)$-frequency undulator gives rise to the new sets of numbers $\{n_i\}$ corresponding to the given $n$ and $m$, viz., they ought to satisfy
\begin{equation}
    \sum_{i=1}^M\la_in_i=n,\qquad\sum_{i=1}^M n_i=m,
\end{equation}
where it has been supposed that the $(M-1)$-frequency undulator is obtained from the $M$-frequency one by turning off the corresponding contribution to the undulator magnetic field: $H^M_{x,y}\rightarrow0$. For the resonance to be strong, the absolute value of the number $n_M$ associated with the activated frequency must not be too large, for example, $n_M=\pm1$.

As the example, we consider the three-frequency undulator with the parameters $(\la_1,\la_2,\la_3)=(2,-3,7)$ and $(K_1,K_2,K_3)=(6.54,8.44,10.85)$, where the undulator strength parameter $K_3$ is enabled from zero to the specified value. The other parameters of this undulator are presented in the caption to Fig. \ref{Plot_Undul_23m7_m_distr}. The radiation from this undulator obeys the selection rule \eqref{sel_rul3-2}, which is written as
\begin{equation}\label{sel_rul3_part}
    m=-2n+5r,\qquad r\in\Z.
\end{equation}
As is seen in Fig. \ref{Plot_Undul_23m7_m_distr}, having turned on the third frequency, there are the modes with certain values of $m$ for all harmonics such that their amplitudes increase significantly. At other values of TAM $m$, the amplitudes of the corresponding modes do not change or decrease. On a qualitative level, the increase of the amplitudes of modes with certain values of $m$ and the spectrum with respect to TAM can be explained by considering the different contributions to the radiation amplitude with given $n$ and $m$ that come from the processes involving the virtual photons characterized by the numbers $(n_1,n_2,n_3)$.

For example, consider the first harmonic, $n=1$. For the two-frequency undulator, i.e., for $K_3=0$, the leading contribution comes from the exchange process mediated by virtual photons which is characterized by the numbers $(n_1,n_2)=(-1,-1)$ and $(n_1,n_2)=(2,1)$  corresponding to $m=-2$ and $m=3$, respectively. The contribution with $m=3$ is less than the contribution with $m=-2$ by three orders of magnitude. On enabling the third frequency, in addition to the existed combinations $(n_1,n_2,n_3)=(-1,-1,0)$, $(n_1,n_2,n_3)=(2,1,0)$, there appear the additional combinations $(n_1,n_2,n_3)=(0,2,1)$, $(n_1,n_2,n_3)=(-3,0,1)$, and $(n_1,n_2,n_3)=(1,-2,-1)$ corresponding to $m=3$, $m=-2$, and $m=-2$, respectively. The presence of these new contributions gives rise approximately to a two-fold increase of the intensity of the mode with $m=-2$, $s=-1$ and the mode with $m=3$, $s=1$.

As for the second harmonic, $n=2$, of the two-frequency undulator, the two dominating modes with definite values of $m$ are described by $(n_1,n_2)=(1,0)$ and $(n_1,n_2)=(-2,-2)$ corresponding to $m=1$ and $m=-4$, respectively. The contribution with $m=-4$ is suppressed approximately by five orders of magnitude as compared to the contribution with $m=1$. Having turned on the third frequency, apart from the combinations $(n_1,n_2,n_3)=(1,0,0)$ and $(n_1,n_2,n_3)=(-2,-2,0)$, there appears the additional combination $(n_1,n_2,n_3)=(-1,1,1)$ yielding a considerable contribution. This contribution corresponds to $m=1$ and results approximately in a three-fold increase of the intensity of the mode with $m=1$, $s=1$. One can verify by changing the phase $\chi_3$ and calculating a change of the intensity of the mode with $m=1$ that it is the combination $(n_1,n_2,n_3)=(-1,1,1)$ which gives the additional contribution to the amplitude with $m=1$.

At the sixth harmonic, $n=6$, there are the two dominating contributions for the two-frequency undulator: $(n_1,n_2)=(0,-2)$ and $(n_1,n_2)=(3,0)$ with $m=-2$ and $m=3$, respectively. Having switched on the third frequency, there appear the additional contributions with $(n_1,n_2,n_3)=(1,1,1)$, $(n_1,n_2,n_3)=(-2,-1,1)$, and $(n_1,n_2,n_3)=(2,-3,-1)$ corresponding to $m=3$, $m=-2$, and $m=-2$. These additional contributions lead approximately to a six-fold increase of the intensities of the mode with $m=3$, $s=1$ and the mode with $m=-2$, $s=-1$.


\begin{figure}[t!]
\centering
\includegraphics*[width=0.98\linewidth]{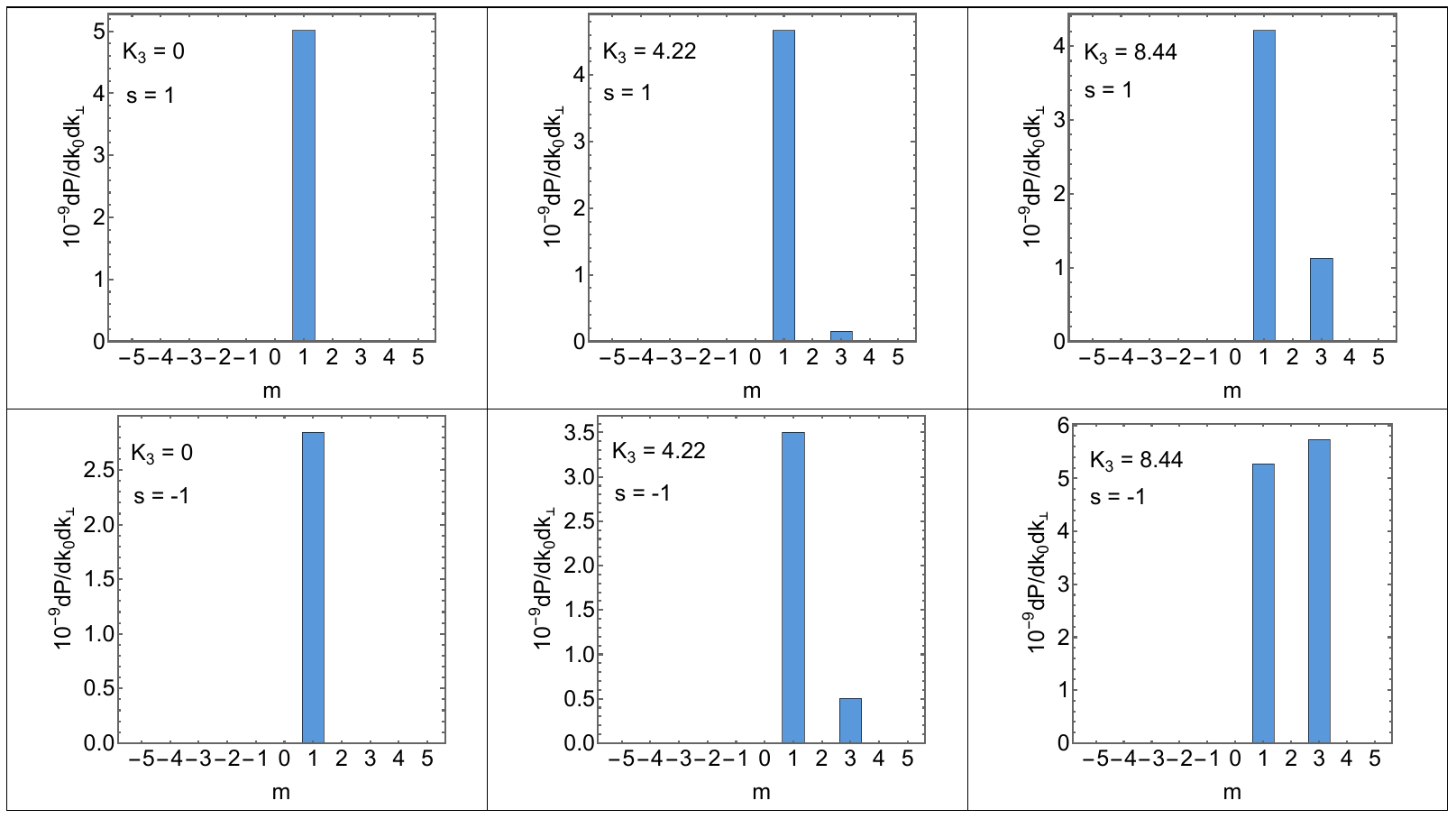}
\caption{{\footnotesize The spectrum of the TAM projections of radiation at the fourth harmonic of the three-frequency helical undulator at $n_\perp =K/\gamma$ and different $K_3$ and $s$. The energy of electrons in the undulator is $8$ GeV, the lengths of sections $(l_1,l_2,l_3)=(1,-3/2,7/2)\times 35$ cm. The magnetic field strengths are $(H^1_{x,y},H^2_{x,y},H^3_{x,y})=(25,1.16\times 10^4,3.22\times 10^3)$ G for the undulator strength parameters $(K_1,K_2,K_3)=(8.18\times 10^{-2},10.85,8.44)$. The energy of photons at the fourth harmonic $k_0=4.58$ eV for these parameters. The undulator phases $\chi_i=0$. Numerical simulations show that, in changing the undulator phases $\{\chi_i\}$, the relative phase between the modes with $m=3$ and $m=1$ agrees with \ref{rel_phase_m31}.}}
\label{Plot_Undul_4_14_5m_m_distr}
\end{figure}


Let us show how, by including the third frequency, one can control the spectrum with respect to $m$ and the relative phases of modes with distinct $m$ of composite states of photons radiated at a given harmonic. As the example, we consider the three-frequency helical undulator with $(\eta_1,\eta_2,\eta_3)=(1,7/2,-5/4)$ and $(K_1,K_2,K_3)=(8.18\times 10^{-2}, 10.85, 8.44)$. These parameters correspond to $(\la_1,\la_2,\la_3)=(4,14,-5)$. The selection rule \eqref{sel_rul3-2} does not impose any restrictions on the values of $m$ at a given $n$. The results of numerical simulations of the radiation intensity for such an undulator are presented in Fig. \ref{Plot_Undul_4_14_5m_m_distr}. We will turn on the third frequency corresponding to $\eta_3=-5/4$. When this frequency is absent, there is the two-frequency undulator with $(\la_1,\la_2)=(2,7)$. The numeration of harmonics is also changed in this case: only even harmonics of the three-frequency undulator are realized in the two-frequency undulator. In particular, the fourth harmonic, $n=4$, of the three-frequency undulator corresponds to the second harmonic, $n=2$, of the two-frequency undulator. The radiation from the two-frequency undulator obeys the selection rule (see \eqref{sel_rul2_n} and formula (61) of the paper \cite{dfund})
\begin{equation}
    m=-2n+5r,\qquad r\in\Z.
\end{equation}
Consider a change of the spectrum with respect to $m$ at the fourth harmonic of the three-frequency undulator in increasing $K_3$. At small $K_3$, the main contribution to the intensity of radiation comes from the combination $(n_1,n_2,n_3)=(1,0,0)$ corresponding to $m=1$. For $K_3=8.44$, there appears the additional contribution $(n'_1,n'_2,n'_3)=(0,2,1)$ corresponding to $m=3$. The relative phase between the modes with $m=3$ and $m=1$ equals
\begin{equation}\label{rel_phase_m31}
    \pm e^{i\tiny\sum_i(n'_i-n_i)\chi_i}=\pm e^{i(2\chi_2+\chi_3-\chi_1)},
\end{equation}
which is confirmed by the numerical simulations.

\section{Conclusion}

Thus our study shows that three-frequency helical undulators can be employed for generation of photons in the composite states with adjustable parameters. The parameters of these states are changed by tuning the parameters of the multifrequency helical undulator -- the amplitudes of the magnetic fields with different periods, $H^i_{x,y}$, and the undulator phases $\chi_i$. In particular, the amplitudes of modes in the composite state of radiated photons can be considerably amplified by enabling a proper additional frequency of the multifrequency undulator with a suitable one-frequency undulator phase $\chi_i$.

We simplified the general formula \eqref{totAmpl2} for the one-particle amplitude of radiation of a twisted photon by a $M$-frequency elliptical undulator. The analog of this formula was derived in \cite{dfund} but it includes much more summations than in formula \eqref{totAmpl2}. Particularizing this formula to the three-frequency case and considering the undulator frequencies with rational ratios, we derived the explicit expression \eqref{totAmpl3} for the amplitude of radiation of a twisted photon at a given radiation harmonic $n$. To this end, we had to solve the linear Diophantine equation \eqref{n_defn} to select those terms in the sum for the radiation amplitude that give the contribution at the given harmonic $n$.

Then we further particularized our investigation and considered the properties of radiation produced by a three-frequency helical undulator, which has the magnetic field \eqref{magnetic_field_3freq_hel}. We obtained a simple expression \eqref{3omegaAmpl2} for the amplitude of radiation of a twisted photon. Thereafter we considered the most interesting case of radiation created by the electron moving near the $z$ axis in the three-frequency helical undulator with rational ratios of frequencies. In this case, the amplitude of radiation of twisted photons obeys the selection rule \eqref{sel_rul3-2} for the TAM projection that says that the radiation spectrum with respect to the TAM projection is equidistant at a given harmonic. By solving the linear Diophantine equation \eqref{sel_rul3-1} expressing the selection rule, we deduced the explicit expression \eqref{amplitude_rad_m} for the amplitude of radiation of a twisted photon and the average number of twisted photons \eqref{average_num_tw_phot} radiated in the three-frequency helical undulator. These expressions allow one to find the relative phases between the modes comprising the composite state of radiated photons. In general, the phases of three modes with different $m$ of a composite state of radiated photon at a given harmonic can be made arbitrary by tuning the undulator phases $\chi_i$. However, these relative phases depend on the other quantum numbers of twisted photons such as the helicity $s$ and $n_\perp=k_\perp/k_0$. This dependence can be eliminated by a special choice of the undulator phases $\chi_i$. One of such choices is simply $\chi_i=\bar\chi$. In this case, the relative phases prove to depend linearly on the TAM projection of the mode at a given harmonic. We also proved in Appendix \ref{Rel_Phases_Modes_App} that this property is valid for the radiation from a two-frequency helical undulator for an arbitrary choice of the undulator phases $\chi_i$.

The numerical simulations presented in Sec. \ref{Num_Simul} corroborate the general theoretical results described above. Moreover, we showed explicitly by changing the parameters of the three-frequency helical undulator that the spectrum with respect to the TAM projections, the amplitudes, and the relative phases of the modes in composite state of photons emitted at a given harmonic can be managed. We also demonstrated how the resonance develops for a suitable choice of the enabling frequency in the multifrequency undulator and its phase.

The multifrequency helical undulators provides a bright and tunable source of photons prepared in composite states with adjustable parameters in a wide range of photon energies. It should be expected that coherent transition, Vavilov-Cherenkov, and undulator radiations from multiperiodic helically microbunched beams with rational ratios of periods possess similar properties. The same seems to be valid for the radiation from charged particles propagating in multicolor laser waves with rational ratios of frequencies. Notice that some properties of radiation of twisted photons by electrons moving in two-color laser waves with multiple frequencies were already investigated in \cite{Taira2018,Jiang2025}. The generalization of the results of the present paper to describe the properties of radiation emitted by beams of particles and to take into account quantum recoil can be done in the same way as given in the papers \cite{BKL4,Bogdanov2019,Bogdanov2020,Takabayashi2025}. We leave the study of these problems for future research.

\paragraph{Acknowledgments.}

The study was supported by the Russian Science Foundation, grant No. 25-21-00283, https://rscf.ru/en/project/25-21-00283/

\appendix
\section{Relative phases of modes}\label{Rel_Phases_Modes_App}

For the relative phases between the contributions with different $m$ in the composite state of radiated photons to be independent of $s$ and $n_\perp$, one can choose the undulator phases $\{\chi_i\}$ in the complex conjugate amplitude \eqref{amplitude_rad_m} such that
\begin{equation}\label{phase_lock}
    \la_{32} \chi_1+\la_{13}\chi_2 + \la_{21}\chi_3 =2\pi \mathrm{d}_{123} r,\qquad r\in\Z.
\end{equation}
In this case, the relative phase between modes with different $m$ is determined by the factor
\begin{equation}
    \pm e^{i\tiny{\sum_i} \bar{n}_i\chi_i}.
\end{equation}
Expressing $\chi_1$ from \eqref{phase_lock} and using the relations \eqref{bar_n_restr}, we obtain for the relative phase
\begin{equation}
    \sum_i \bar{n}_i\chi_i=\frac{\chi_{3}-\chi_2}{\la_{32}}n+\frac{2\pi  \mathrm{d}_{123}r\bar{n}_1+(\chi_2\la_3-\chi_3\la_2)m}{\la_{32}}\rightarrow \frac{\chi_2 \la_3-\chi_3 \la_2 -2\pi r \mathrm{d}_{23} b^0_1}{\la_{32}} m,
\end{equation}
where, after the arrow, $\bar{n}_1$ has been substituted from \eqref{bar_n_i} into the expression and the terms independent of $m$ have been discarded. We see that, in the case under consideration, the relative phase of modes with different $m$ is proportional to $m$ as in the case of $\chi_i=\bar{\chi}$.

Notice that for a two-frequency helical undulator studied in \cite{dfund} the same property holds for an arbitrary choice of phases $(\chi_1,\chi_2)$. Indeed, according to formulas (54), (56), and (61) of \cite{dfund}, the numbers of virtual photons $(n_1,n_2)$ obey the relations
\begin{equation}\label{sel_rul2}
    \la_1 n_1+\la_2 n_2=n,\qquad n_1+n_2=m.
\end{equation}
The relative phase of modes in a composite state is determined by
\begin{equation}
    \pm e^{i(n_1\chi_1+n_2\chi_2)}.
\end{equation}
Equations \eqref{sel_rul2} have a solution when
\begin{equation}\label{sel_rul2_n}
    m=n(n_1^0+n_2^0)+\la_{21}k,\qquad k\in\Z,
\end{equation}
where $(n_1^0,n_2^0)$ are the B\'{e}zout coefficients for $(\la_1,\la_2)$ and
\begin{equation}
    n_1=nn_1^0+\la_2 k,\qquad n_2=nn_2^0-\la_1 k.
\end{equation}
Therefore, the relative phase of modes in a composite state is given by
\begin{equation}
    n_1\chi_1+n_2\chi_2=\frac{\chi_2-\chi_1}{\la_{21}}n +\frac{\chi_1\la_2-\chi_2\la_1}{\la_{21}}m
\end{equation}
and depends linearly on $m$ for a fixed $n$.

\end{document}